\documentclass[sigconf]{acmart}
\makeatletter

\let\@authorsaddresses\@empty

\makeatother

\acmConference[arxiv]{arxiv pre-print versions}{2023}{}
\renewcommand\footnotetextcopyrightpermission[1]{}

\AtBeginDocument{%
  }

\usepackage{amsmath, amsfonts}
\usepackage{algorithmic}
\usepackage{graphicx}
\usepackage{textcomp}
\usepackage{xcolor}
\usepackage{listings,listings-rust}
\usepackage{caption,subcaption}
\usepackage{enumerate}
\usepackage{multirow}
\usepackage{color}
\usepackage{colortbl}
\usepackage[ruled,vlined]{algorithm2e}
\usepackage{amsfonts}
\usepackage{threeparttable}
\usepackage{booktabs}
\usepackage{makecell}
\usepackage{cleveref}
\usepackage{soul}

\definecolor{gray_blank}{gray}{.7}
\definecolor{gray_item}{gray}{.7}

\definecolor{generic}{rgb}{0.75, 0, 0}
\definecolor{mut}{rgb}{0, 0.5, 0}
\definecolor{primitive}{rgb}{0, 0, 0.5}

\newtheorem{goal}{\textbf{GOAL}}[section]
\newtheorem{filter}{\textbf{FILTER}}[section]

\crefname{goal}{Goal}{Goal}

\begin{document}

\title{Is \textsc{unsafe} an Achilles' Heel? A Comprehensive Study of Safety Requirements in Unsafe Rust Programming}

\author{Mohan Cui}
\affiliation{%
  \institution{School of Computer Science, \\Fudan University}
  \country{}
}

\author{Suran Sun}
\affiliation{%
  \institution{School of Computer Science, \\Fudan University}
  \country{}
}

\author{Hui Xu}
\authornote{Corresponding author.}
\affiliation{%
  \institution{School of Computer Science, \\Fudan University}
  \country{}
}

\author{Yangfan Zhou}
\affiliation{%
  \institution{School of Computer Science, \\Fudan University}
  \country{}
}

\renewcommand{\shortauthors}{Cui et al.}

\begin{abstract}
Rust is an emerging, strongly-typed programming language focusing on efficiency and memory safety. With increasing projects adopting Rust, knowing how to use Unsafe Rust is crucial for Rust security. We observed that the description of safety requirements needs to be unified in Unsafe Rust programming. Current unsafe API documents in the standard library exhibited variations, including inconsistency and insufficiency. To enhance Rust security, we suggest unsafe API documents to list systematic descriptions of safety requirements for users to follow.

In this paper, we conducted the first comprehensive empirical study on safety requirements across unsafe boundaries. We studied unsafe API documents in the standard library and defined 19 safety properties (SP). We then completed the data labeling on 416 unsafe APIs while analyzing their correlation to find interpretable results. To validate the practical usability and SP coverage, we categorized existing Rust CVEs until 2023-07-08 and performed a statistical analysis of std unsafe API usage toward the crates.io ecosystem. In addition, we conducted a user survey to gain insights into four aspects from experienced Rust programmers. We finally received 50 valid responses and confirmed our classification with statistical significance.
\end{abstract}



\maketitle

\section{Introduction}
Rust is an emerging system programming language focusing on memory safety and efficiency~\cite{matsakis2014rust}. It provides memory-safe guarantees via compile-time checks~\cite{fulton2021benefits}; consequently, programmers must adhere to various syntactic constraints to satisfy the verification~\cite{zhu2022learning}. As a system programming language, it employs several zero-cost abstractions~\cite{jung2017rustbelt} to transform data without sacrificing performance~\cite{klabnik2019rust} (\textit{e.g.,} generic types). Although Rust has a steep learning curve~\cite{fulton2021benefits}, it has attracted many programmers due to its safety and efficiency~\cite{popular}. Since 2016, Rust has been the most popular programming language in the open-source community~\cite{survey16,survey17,survey18,survey19,survey20,survey21,survey22}, with many projects refactoring code in Rust~\cite{shen2020occlum,levy2017tock,levy2017multiprogramming}.

\begin{figure}[]
\begin{subfigure}[]{0.45\textwidth}
\begin{lstlisting}[language=Rust, style=colouredRust, label=list:take, caption=Source code of ManuallyDrop::take in Rust std.\\]
// impl<T> ManuallyDrop<T>
pub unsafe fn take(slot: &mut ManuallyDrop<T>) -> T {
    // SAFETY: we are reading from a reference, which is
    // guaranteed to be valid for reads.
    unsafe { ptr::read(&slot.value) }
}
\end{lstlisting}
\label{fig:relation1}
\end{subfigure}

\begin{subfigure}[]{0.45\textwidth}
\begin{lstlisting}[language=c, style=colouredRust, label=list:takedoc, caption=Document of ManuallyDrop::take in Rust 1.70.\\]
Takes the value from the ManuallyDrop<T> container out.
This method is primarily intended for moving out values in drop. Instead of using ManuallyDrop::drop to manually drop the value, you can use this method to take the value and use it however desired.
Whenever possible, it is preferable to use into_inner instead, which prevents duplicating the content of the ManuallyDrop<T>.
Safety
This function semantically moves out the contained value without preventing further usage, leaving the state of this container unchanged. It is your responsibility to ensure that this ManuallyDrop is not used again.
\end{lstlisting}
\end{subfigure}
\caption{Example of an unsafe API in Rust std. Listing~\ref{list:take} provides the source code of an unsafe method of struct ManuallyDrop<T>. Listing~\ref{list:takedoc} extracts the document in Rust 1.70. The document introduces the usage, functionality, and safety requirements to comply with by using a section labeled \textit{Safety}.}
\label{fig:take}
\end{figure}

\begin{table*}[]
\caption{A list of APIs with similar side effects and their document slices in Rust 1.70. These APIs accept mutable pointers as input and return a typed owner. In previous research, it has been reported that misuse of these APIs may result in double-free issues. The consistency and clarity of these documents need improvement: Related descriptions are not always located within the \textit{Safety} section, and the description of the safety requirement (underlined) or side effect (bolded) is insufficient. We expect a specific error type to be defined.}
\label{table:unsafe_ctors}
\resizebox{\linewidth}{!}{
\begin{tabular}{ccp{11cm}c}
    \toprule[1pt]
	\textbf{Implemented Type}    & \textbf{Unsafe Method} & \textbf{Safety Description Slices in API Documents.} \\
	\midrule[1pt]
	\multirow{3}{*}{\textbf{impl}<\textcolor{generic}{\textbf{\textcolor{generic}{\textbf{T: ?Sized}}}}> \textcolor{mut}{*mut} \textcolor{generic}{\textbf{T}}} & \multirow{3}{*}{\textbf{fn} \textbf{read}(\textbf{self}) -> \textcolor{generic}{\textbf{T}}} & read creates a bitwise copy of T, regardless of whether T is Copy. \ul{If T is not Copy, using both the returned value and the value at *src can \textbf{violate memory safety}.} Note that assigning to *src counts as a use because it will attempt to drop the value at *src.\\
	\midrule[0.5pt]
	\multirow{3}{*}{\textbf{impl}<\textcolor{generic}{\textbf{T}}> \textbf{ManuallyDrop}<\textcolor{generic}{\textbf{T}}>} & \multirow{3}{*}{\textbf{fn} \textbf{take}(\textcolor{mut}{\&mut} \textbf{ManuallyDrop}<\textcolor{generic}{\textbf{T}}>) -> \textcolor{generic}{\textbf{T}}} & This function semantically moves out the contained value without preventing further usage, leaving the state of this container unchanged. It is your responsibility to \ul{ensure that this ManuallyDrop is not used again.} \\
	\midrule[0.5pt]
	\multirow{3}{*}{\textbf{impl}<\textcolor{generic}{\textbf{T: ?Sized}}> \textbf{Box}<\textcolor{generic}{\textbf{T}}>} & \multirow{3}{*}{\textbf{fn} \textbf{from\_raw}(\textcolor{mut}{*mut} \textcolor{generic}{\textbf{T}}) -> \textbf{Self}} & This function is unsafe because improper use may lead to \textbf{memory problems}. For example, \ul{a \textbf{double-free} may occur if the function is called twice on the same raw pointer.}\\
	\midrule[0.5pt]
	\multirow{3}{*}{\textbf{impl}<\textcolor{generic}{\textbf{\textcolor{generic}{\textbf{T: ?Sized}}}}> \textbf{Rc}<\textcolor{generic}{\textbf{T}}>} & \multirow{3}{*}{\textbf{fn} \textbf{from\_raw}(\textcolor{mut}{*const} \textcolor{generic}{\textbf{T}}) -> \textbf{Self}} & \ul{The raw pointer must have been previously returned by a call to Rc<U>::into\_raw where U must have the same size and alignment as T.} The user of from\_raw has to \ul{make sure a specific value of T is only dropped once.} \\
	\midrule[0.5pt]
	\multirow{3}{*}{\textbf{impl} \textbf{CString}} & \multirow{3}{*}{\textbf{fn} \textbf{from\_raw}(\textcolor{mut}{*mut} \textcolor{primitive}{\textbf{c\_char}}) -> \textbf{Self}}  & This should only ever be called with \ul{a pointer that was earlier obtained by calling CString::into\_raw.} Other usage (e.g., \ul{trying to take ownership of a string that was allocated by foreign code}) is likely to lead to \textbf{undefined behavior} or \textbf{allocator corruption}.\\
	\midrule[0.5pt]
	\multirow{3}{*}{\textbf{impl}<\textcolor{generic}{\textbf{T}}> \textbf{Vec}<\textcolor{generic}{\textbf{T}}>} & \multirow{3}{*}{\textbf{fn} \textbf{from\_raw\_parts}(\textcolor{mut}{*mut} \textcolor{generic}{\textbf{T}}, \textcolor{primitive}{\textbf{usize}}, \textcolor{primitive}{\textbf{usize}}) -> \textbf{Self}} & The ownership of ptr is effectively transferred to the Vec<T> which may then deallocate, reallocate or change the contents of memory pointed to by the pointer at will. \ul{Ensure that nothing else uses the pointer after calling this function.}\\
    \midrule[0.5pt]
    \multirow{3}{*}{\textbf{impl} \textbf{String}} & \multirow{3}{*}{\textbf{fn} \textbf{from\_raw\_parts}(\textcolor{mut}{*mut} \textcolor{primitive}{\textbf{u8}}, \textcolor{primitive}{\textbf{usize}}, \textcolor{primitive}{\textbf{usize}}) -> \textbf{Self}} & The ownership of buf is effectively transferred to the String which may then deallocate, reallocate or change the contents of memory pointed to by the pointer at will. \ul{Ensure that nothing else uses the pointer after calling this function.}\\
    \bottomrule[1pt]
\end{tabular}
}
\end{table*}

As Rust continues to evolve, knowing how to use Unsafe Rust is essential for Rust security~\cite{astrauskas2020programmers}. Safety isolation is one of the revolutionary innovations introduced by Rust~\cite{qin2020understanding}. It divides the portions the compiler can ensure safety into Safe Rust and adds the \textit{unsafe} keyword as the superset~\cite{klabnik2019rust}. The primary document defines Unsafe Rust as a keyword and a set of operations~\cite{unsafesyntactic}. Any code with unsafe operations must be wrapped in an unsafe block. If not, programmers will trigger compilation errors. Without strict compiler checks in the unsafe scope, Rust developers may become insensitive to satisfying safety requirements, which is error-prone to causing undefined behavior (UB).

How does Rust document safety requirements for unsafe operations? We observed that most safety requirements are specified on a \textbf{\textit{Safety}} label in Rust std. The Rust standard library~\cite{ruststd} provides documents for unsafe APIs that are relatively comprehensive. As shown in Figure~\ref{fig:take}, we chose one API as the typical example. When calling \texttt{ManuallyDrop::take}~\cite{take}, it has a safety requirement to be manually reviewed: Users cannot use this container again. Otherwise, it would trigger undefined behavior. Its implementation calls unsafe function \texttt{ptr::read}~\cite{read} (line 5), prompting us to think they may have analogous safety descriptions.

Unfortunately, Unsafe Rust does not provide developers with unified safety descriptions or systemic safety requirements. Recent research found that misusing several unsafe APIs may result in memory-safety issues~\cite{xu2021memory}, where overlapped owners can be created and cause double free~\cite{cui2023safedrop, bae2021rudra}, such as \texttt{ManuallyDrop::take} and \texttt{*mut T::read}. Table~\ref{table:unsafe_ctors} lists them with documents in Rust 1.70. Like \texttt{ManuallyDrop::take}, using a \textbf{\textit{Safety}} label to start the safety description is intuitive. The majority of the listed APIs adhere to this criterion, such as implementations for \texttt{String}, \texttt{Vec<T>}, \texttt{CString}, and \texttt{Box<T>}. However, the \texttt{Rc<T>} lacks the \textbf{\textit{Safety}} section, and the related issue caused by \texttt{read} is described in the outer section. At last, the texts of side effects exhibit differences: Only \texttt{Box<T>} explicitly states the potential double-free that may arise.

The unsafe API documents should systematically classify safety requirements for users to comply with. This paper comprehensively categorizes fine-grained safety requirements when crossing unsafe boundaries. In general, this paper seeks to address the following research questions (RQs):

\begin{enumerate}[0]
\item[$\bullet$] \noindent \textbf{RQ-1.} What finer-grained safety properties (requirements) should be satisfied across the Unsafe Rust boundary? (\S\ref{sec:sp})
\item[$\bullet$] \noindent \textbf{RQ-2.} Can those safety properties cover existing vulnerabilities caused by Unsafe Rust? (\S\ref{sec:cveio})
\item[$\bullet$] \noindent \textbf{RQ-3.} How helpful are those safety properties for real-world Unsafe Rust programming? (\S\ref{sec:survey})
\end{enumerate}

For each RQ, our study introduces several sub-experiments. To answer RQ1, we extracted all public unsafe APIs within the Rust standard library~\cite{ruststd} and manually audited the document. We categorized the safety requirements across the unsafe boundary as \textbf{Safety Properties} (SPs). We completed the data labeling for those APIs and then conducted a correlation analysis to find interpretable results. To answer RQ2, we examined all Rust CVEs~\cite{cve} until 2023-07-08 and filtered through the root causes by misusing unsafe code, categorizing them according to Safety Properties to validate our classification. Then, we collected and analyzed the distribution of unsafe APIs within the crates.io~\cite{cratesio} ecosystem. To answer RQ3, we surveyed experienced Rust developers. We provided participants with the definition of each SP and its minimal Proof of Concept (PoC). We studied whether the developers acknowledged our categorization and whether each Safety Property was beneficial for unsafe programming.

Reviewing documents of unsafe APIs in Rust std, we performed an audit on 416 unsafe APIs. As a result, we identified and defined 19 safety properties (SP), categorized into two major categories: precondition and postcondition. Subsequently, we completed the SP labeling for all unsafe APIs, creating two datasets for correlation analysis. The results revealed six crucial SPs that users need to satisfy when dereferencing. Next, we classified the existing Rust CVEs based on safety properties, with 196 of 404 resulting from unsafe code. Notably, 86.73\% of these errors were attributable to misuse of the standard library. Therefore, we conducted a statistical analysis of std unsafe API usage for the Rust ecosystem, which included 103,516 libraries on crates.io. Finally, we conducted user surveys targeting developers with over one year of Rust experience, having written over 5,000 lines of code and using over 1,000 lines of unsafe code. The evaluations for each SP were rated on four dimensions: precision, significance, usability, and frequency. We received 50 valid responses and conducted data analysis on them.

Our main contributions are listed as follows:

\begin{enumerate}[0]
\item[$\bullet$] \noindent We performed the first empirical study by learning unsafe API documents from the standard library to classify safety requirements across unsafe Rust boundaries.
\item[$\bullet$] \noindent We classified 19 safety properties into two categories. All std-unsafe APIs were audited and labeled with safety properties. The labeled data were evaluated via correlation analysis, yielding interpretable results.
\item[$\bullet$] \noindent We categorized all Rust CVEs based on safety properties, forming a collection of related issues that can serve as a benchmark. Unsafe API usage statistics are collected within crates.io to understand the usage frequency of unsafe APIs.
\item[$\bullet$] \noindent We conducted an online survey and confirmed our categorization of safety properties with statistical significance.
\end{enumerate}

\section{Background}

\subsection{Working with Unsafe Rust}
Rust is subdivided into Safe Rust and Unsafe Rust, with Unsafe Rust being a superset~\cite{qin2020understanding}. Safe Rust ensures type and memory safety, preventing undefined behavior~\cite{astrauskas2020programmers}. However, it lacks low-level controls over implementation details (\textit{e.g.,} manual memory management). Unsafe Rust is an essential design feature to achieve low-level control at the system level~\cite{rustonomicon}. It is employed if it has performance requirements or needs to interact with operating systems, hardware, or other programming languages.

\noindent \textbf{\textit{unsafe} Keyword}. \textit{unsafe} keyword can be used in declarations and code blocks. The first scenario indicates that the functions cannot be called in the safe code. Misuse may trigger undefined behavior. In code blocks, it signifies the scope that may violate safety guarantees without compiler-time checks, and the code requires manual auditing to ensure safety. This keyword acts as a railing, separating the safe and unsafe portions: All unsafe parts are encapsulated within this scope. The trust relationship between safe and unsafe parts is asymmetric~\cite{rustonomicon}. When using an unsafe block, careful inspection is required to ensure that the data from the safe portion adheres to the contracts of the unsafe APIs. Conversely, when writing safe code, it is assumed that the unsafe code is correct and would not trigger undefined behavior.

\noindent \textbf{\textit{unsafe} Operations}. Safe Rust and Unsafe Rust are designed for different scenarios. Safe Rust is a safe programming language designed for tasks that do not require low-level interactions. Contrariwise, Unsafe Rust fully leverages the capabilities of a systems-level programming language. Unlike languages such as C/C++, which are inherently unsafe, Unsafe Rust still requires adherence to certain contracts from the safe portion, such as ownership. The main differences in Unsafe Rust are that you can \textit{1) Dereference raw pointers; 2) Call unsafe functions; 3) Implement unsafe traits; 4) Mutate static variables; and 5) Access fields of unions}~\cite{rustonomicon}. These operations provide flexibility but come with the responsibility of the users to manually ensure correctness and safety.

Recent empirical research~\cite{zhu2022learning} suggests that Rust's safety mechanisms could be more learner-friendly. This study explored the learning challenges of its safety mechanisms by analyzing Stack Overflow comments and conducting user surveys, but it is restricted to Safe Rust. Instead, learning and utilizing Unsafe Rust is a prerequisite for advanced Rust developers. 

\subsection{Undefined Behavior in Rust}\label{sec:ub}
The undefined behavior in Rust is limited to include~\cite{reference}:

\begin{enumerate}[0]
\item[$\bullet$] \noindent Dereferencing (using the \texttt{*} operator on) dangling or unaligned raw pointers.
\item[$\bullet$] \noindent Breaking the pointer aliasing rules. References and boxes must not be dangling while they are alive.
\item[$\bullet$] \noindent Calling a function with the wrong call ABI or unwinding from a function with the wrong unwind ABI.
\item[$\bullet$] \noindent Executing code compiled with platform features that the current thread of execution does not support.
\item[$\bullet$] \noindent Producing invalid values, as explained in Table~\ref{table:ub}, even in private fields and locals.
\item[$\bullet$] \noindent Mutating immutable data. All data inside a const item, reached through a shared reference or owned by an immutable binding, is immutable.
\item[$\bullet$] \noindent Causing data races.
\item[$\bullet$] \noindent Invoking undefined behavior via compiler intrinsics.
\item[$\bullet$] \noindent Incorrect use of inline assembly.
\end{enumerate}

This categorization is based on the side effects introduced by unsafe code. Since no formal model of Rust's semantics defines precisely what is and is not permitted in unsafe code~\cite{reference}, additional behavior may be deemed vulnerable. In this paper, we additionally introduce the following issues as program vulnerabilities if they are triggered by unsafe code:

\begin{enumerate}[0]
\item[$\bullet$] \noindent \underline{Causing a memory leak and exiting without calling destructors.}
\item[$\bullet$] \noindent \underline{Triggering an unreachable path then aborting (or panicking).}
\item[$\bullet$] \noindent \underline{Arithmetic overflow.}
\end{enumerate}

These undefined behavior and vulnerabilities serve as the basis for classifying safety requirements. Other errors fall outside the scope (\textit{e.g.,} deadlocks and logic errors).

\begin{table}[]
\caption{Invalid value for Rust types, alone or as a field of a compound type, will trigger undefined behavior.}
\label{table:ub}
\resizebox{\linewidth}{!}{
\begin{tabular}{cp{7cm}c}
    \toprule[1pt]
	\textbf{Rust Type} & \textbf{Invalid Value}\\
    \midrule[1pt]
    \textbf{!} & Invalid for all values.\\
	\hline
	\textbf{\texttt{bool}} & Not \texttt{0} or \texttt{1} in bytes.\\
	\hline
	\textbf{\texttt{char}} & Outside \texttt{[0x0, 0xD7FF]} \& \texttt{[0xE000, 0x10FFFF]}.\\
	\hline
	\textbf{\texttt{str}} & Has uninitialized memory.\\
	\hline
	\textbf{numeric \texttt{i*/u*/f*}} & Reads from uninitialized memory.\\
	\hline
	\textbf{\texttt{enum}} & Has an invalid discriminant.\\
	\hline
	\textbf{reference} & Dangling, unaligned, or pointing to an invalid value.\\
	\hline
	\textbf{raw pointer} & Reads from uninitialized memory.\\
	\hline
	\textbf{\texttt{Box}} & Dangling, unaligned, or pointing to an invalid value.\\
	\hline
	\textbf{\texttt{fn} pointer} & NULL. \\
	\hline
	\multirow{4}{*}{\textbf{wide reference}} & Has invalid metadata. \texttt{dyn} \texttt{Trait} is invalid if it is not a pointer to a vtable for \texttt{Trait} that matches the actual dynamic trait the pointer or reference points to, and \texttt{slice} is invalid if the length is not a valid \texttt{usize}. \\
	\hline
	\textbf{custom type} & Has one of those custom invalid values. \\
    \bottomrule[1pt]
\end{tabular}
}
\end{table}

\section{Studying Unsafe Documents in STD}\label{sec:sp}
This section presents how we extract and define systematic safety requirements as \textbf{Safety Properties} (SP) from the existing unsafe documents in the standard library~\cite{ruststd}. Our classification allows us to clarify the primary conditions and constraints necessary for Unsafe Rust, hence answering RQ1.

\begin{table*}[]
\caption{Safety properties learned from unsafe API documents in Rust standard library. Precondition and postcondition are two primary categories, and they have 19 subitems in total. We present the sum of the labeled unsafe API for each safety property and provide a detailed definition. Each safety requirement was also given a typical unsafe API as an example. Note that each SP may contain various sub-scenarios.}
\label{table:sp}
\resizebox{\linewidth}{!}{
\begin{threeparttable}
\begin{tabular}{|c|c|p{11.5cm}|c|c}
    \toprule[1pt]
	\textbf{Safety Property (SP) } & \textbf{SUM} & \textbf{Definition and the safety requirement of each Safety Property.} & \textbf{Unsafe API Example} \\	
	\hline
	\rowcolor{gray_blank} \multicolumn{4}{|c|}{}\\
	\rowcolor{gray_blank} \multicolumn{4}{|c|}{\multirow{-2}{*}{\textbf{Precondition Safety Property}}}\\
	\hline
	\multirow{2}{*}{\textbf{Const-Numeric Bound}} & \multirow{2}{*}{\textbf{72}} & Relational operations allow for \textbf{compile-time} determination of the \textbf{constant} numerical boundaries on one side of an expression, including overflow check, index check, etc. & \multirow{2}{*}{\textbf{impl}<\textbf{\textcolor{generic}{\textbf{T: ?Sized}}}> \textcolor{mut}{*mut} \textbf{\textcolor{generic}{\textbf{T}}}::offset\_from}\\
	\hline
	\multirow{2}{*}{\textbf{Relative-Numeric Bound}} & \multirow{2}{*}{\textbf{114}} & Relational operations involve expressions where \textbf{neither side} is a constant numeric, including address boundary check, overlap check, size check, variable comparison, etc. & \multirow{2}{*}{\textbf{trait} \textbf{Allocator}::grow}\\
	\hline
	\multirow{2}{*}{\textbf{Encoding}} & \multirow{2}{*}{\textbf{16}} & Encoding format of the string, includes valid \textbf{UTF-8} string, valid \textbf{ASCII} string (in bytes), and valid \textbf{C-compatible} string (nul-terminated trailing with no \texttt{nul} bytes in the middle). & \multirow{2}{*}{\textbf{impl} \textbf{String}::from\_utf8\_unchecked}\\
	\hline
	\multirow{2}{*}{\textbf{Allocated}} & \multirow{2}{*}{\textbf{134}} & Value stored in the \textbf{allocated memory}, including data in the valid stack frame and allocated heap chunk, which cannot be \texttt{NULL} or dangling. & \multirow{2}{*}{\textbf{impl}<\textbf{\textcolor{generic}{\textbf{T: Sized}}}> \textbf{NonNull}<\textbf{\textcolor{generic}{\textbf{T}}}>::new\_unchecked}\\
	\hline
	\multirow{2}{*}{\textbf{Initialized}} & \multirow{2}{*}{\textbf{59}} & Value that has been initialized can be divided into two scenarios: \textbf{fully initialized} and \textbf{partially initialized}. The initialized value must be \textbf{valid} at the given type (a.k.a. \textbf{typed}). & \multirow{2}{*}{\textbf{impl}<\textbf{\textcolor{generic}{\textbf{T}}}> \textbf{MaybeUninit}<\textbf{\textcolor{generic}{\textbf{T}}}>::assume\_init}\\
	\hline
	\multirow{2}{*}{\textbf{Dereferencable}} & \multirow{2}{*}{\textbf{96}} & The memory range of the given size starting at the pointer must all be within the bounds of a \textbf{single allocated object}. & \multirow{2}{*}{\textbf{impl}<\textbf{\textcolor{generic}{\textbf{T: ?Sized}}}> \textcolor{mut}{*const} \textbf{\textcolor{generic}{\textbf{T}}}::as\_ref}\\
	\hline
	\multirow{2}{*}{\textbf{Aligned}} & \multirow{2}{*}{\textbf{67}} & Value is \textbf{properly aligned} via a specific \textbf{allocator} or the attribute \texttt{\#[repr]}, including the alignment and the padding of one Rust type. & \multirow{2}{*}{\textbf{impl}<\textbf{\textcolor{generic}{\textbf{T: ?Sized}}}> \textcolor{mut}{*mut} \textbf{\textcolor{generic}{\textbf{T}}}::swap}\\
	\hline
	\multirow{4}{*}{\textbf{Consistent Layout}} & \multirow{4}{*}{\textbf{110}} & Restriction on Type Layout, including 1) The \textbf{pointer's} type must be compatible with the \textbf{pointee's} type; 2) The \textbf{contained value} must be compatible with the generic parameter for the smart pointer; and 3) Two types are \textbf{safely transmutable}: The bits of one type can be \textbf{reinterpreted} as another type (bitwise move safely of one type into another). & \multirow{4}{*}{\textbf{impl}<\textbf{\textcolor{generic}{\textbf{T: ?Sized}}}> \textcolor{mut}{*mut} \textbf{\textcolor{generic}{\textbf{T}}}::read}\\
	\hline
	\multirow{2}{*}{\textbf{Unreachable}} & \multirow{2}{*}{\textbf{9}} & Specific value will trigger \textbf{unreachable data flow}, such as \textbf{enumeration} index (\textbf{variance}), boolean value, closure and etc.  & \multirow{2}{*}{\textbf{impl}<\textbf{\textcolor{generic}{\textbf{T}}}> \textbf{Option}<\textbf{\textcolor{generic}{\textbf{T}}}>::unwrap\_unchecked}\\
	\hline
	\multirow{3}{*}{\textbf{Exotically Sized Type}} & \multirow{3}{*}{\textbf{24}} & Restrictions on Exotically Sized Types (EST), including \textbf{Dynamically Sized Types} (DST) that lack a statically known size, such as trait objects and slices; \textbf{Zero Sized Types} (ZST) that occupy no space. & \multirow{3}{*}{\textbf{trait} \textbf{GlobalAlloc}::alloc}\\
	\hline
	\multirow{2}{*}{\textbf{System IO}} & \multirow{2}{*}{\textbf{25}} & Variables related to the \textbf{system IO} depends on the \textbf{target platform}, including TCP sockets, handles, and file descriptors. & \multirow{2}{*}{\textbf{trait} \textbf{FromRawFd}::from\_raw\_fd}\\
	\hline
	\multirow{2}{*}{\textbf{Thread}} & \multirow{2}{*}{\textbf{3}} & Types that can be \textbf{transferred} across threads (\textbf{Send}) or types that can be safe to \textbf{share references} between threads (\textbf{Sync}), respectively. & \multirow{2}{*}{\textbf{std}::\textbf{marker}::\textbf{Sync}}\\
	\hline
	\rowcolor{gray_blank} \multicolumn{4}{|c|}{}\\
	\rowcolor{gray_blank} \multicolumn{4}{|c|}{\multirow{-2}{*}{\textbf{Postcondition Safety Property}}}\\
	\hline
	\multirow{2}{*}{\textbf{Dual Owner}} & \multirow{2}{*}{\textbf{31}} & Multiple owners (\textbf{overlapped objects}) that share the same memory in the ownership system by \textbf{retaking the owner} or \textbf{creating a bitwise copy}.  & \multirow{2}{*}{\textbf{impl}<{\textcolor{generic}{\textbf{T: ?Sized}}}> \textbf{Box}<\textcolor{generic}{\textbf{T}}>::from\_raw}\\
	\hline
	\multirow{4}{*}{\textbf{Aliasing \& Mutating}} & \multirow{4}{*}{\textbf{30}} & Aliasing and mutating rules may be violated, including 1) The presence of \textbf{multiple} mutable references; 2) The \textbf{simultaneous} presence of mutable and shared references, and the memory the pointer points to cannot get mutated (\textbf{frozen}); 3) Mutating \textbf{immutable data} owned by an immutable binding. & \multirow{4}{*}{\textbf{impl} \textbf{CStr}::from\_ptr}\\
	\hline
	\multirow{2}{*}{\textbf{Outliving}} & \multirow{2}{*}{\textbf{28}} & \textbf{Arbitrary lifetime} (unbounded) that becomes as big as context demands or \textbf{spawned thread}, may outlive the pointed memory. & \multirow{2}{*}{\textbf{impl}<\textbf{\textcolor{generic}{\textbf{T: ?Sized}}}> \textcolor{mut}{*const} \textcolor{generic}{\textbf{T}}::as\_uninit\_ref}\\
	\hline
	\multirow{2}{*}{\textbf{Untyped}} & \multirow{2}{*}{\textbf{20}} & Value may not be in the \textbf{initialized state}, or the \textbf{byte pattern} represents an \textbf{invalid value} of its type. & \multirow{2}{*}{\textbf{core}::\textbf{mem}::zeroed}\\
	\hline
	\textbf{Freed} & \textbf{17} & Value may be manually \textbf{freed} or \textbf{released} by automated \textbf{\texttt{drop()}} instruction. & \textbf{impl}<\textbf{\textcolor{generic}{\textbf{T: ?Sized}}}> \textbf{ManuallyDrop}<\textcolor{generic}{\textbf{T}}>::drop\\
	\hline
	\textbf{Leaked} & \textbf{13} & Value may be \textbf{leaked} or \textbf{escaped} from the ownership system. & \textbf{impl}<\textbf{\textcolor{generic}{\textbf{T: ?Sized}}}> \textcolor{mut}{*mut} \textcolor{generic}{\textbf{T}}::write\\
	\hline
	\textbf{Pinned} & \textbf{5} & Value may be \textbf{moved}, although it ought to be pinned. & \textbf{impl}<\textbf{\textcolor{generic}{\textbf{P: Deref}}}> \textbf{Pin}<\textcolor{generic}{\textbf{P}}>::new\_unchecked\\
    \bottomrule[1pt]
\end{tabular}
\begin{tablenotes}
	\footnotesize
	\item[1] \texttt{Send}~\cite{send} and \texttt{Sync}~\cite{sync} are unsafe traits (markers) that are automatically implemented by the compiler when it determines that they are required. Therefore, they lack associated methods.
	\item[2] The difference between \texttt{DualOwner} and \texttt{AliasingMutating} is that \texttt{DualOwner} only focuses on objects instead of pointers and references.
\end{tablenotes}
\end{threeparttable}
}
\end{table*}

\subsection{Preprocess on Rust Documents}
\texttt{Rustdoc}~\cite{rustdoc} is the document system for the Rust programs, which enables the description of functionalities, requirements, expected results, and sample code snippets for APIs and crates. Intuitively, input requirements and side effects within an unsafe API must be explicitly specified in \texttt{Rustdoc}. We found that the document in the standard library is one of the most comprehensive resources for safety annotations within the Rust ecosystem. We thus audited documents of all public unsafe methods within the standard library as the knowledge base.

\subsubsection{Design Goals} Table~\ref{table:unsafe_ctors} reveals that even in the standard library: (i) the expression of the same safety requirement is not universally consistent; (ii) the enumeration of the safety requirements and side effects is not always sufficient. Thus, we manually categorize and define a series of finer-grained safety requirements as \textbf{Safety Properties} (SP), which need to satisfy the following design goals:

\begin{figure*}[]
\begin{subfigure}[]{0.49\textwidth}
	\includegraphics[width=\textwidth]{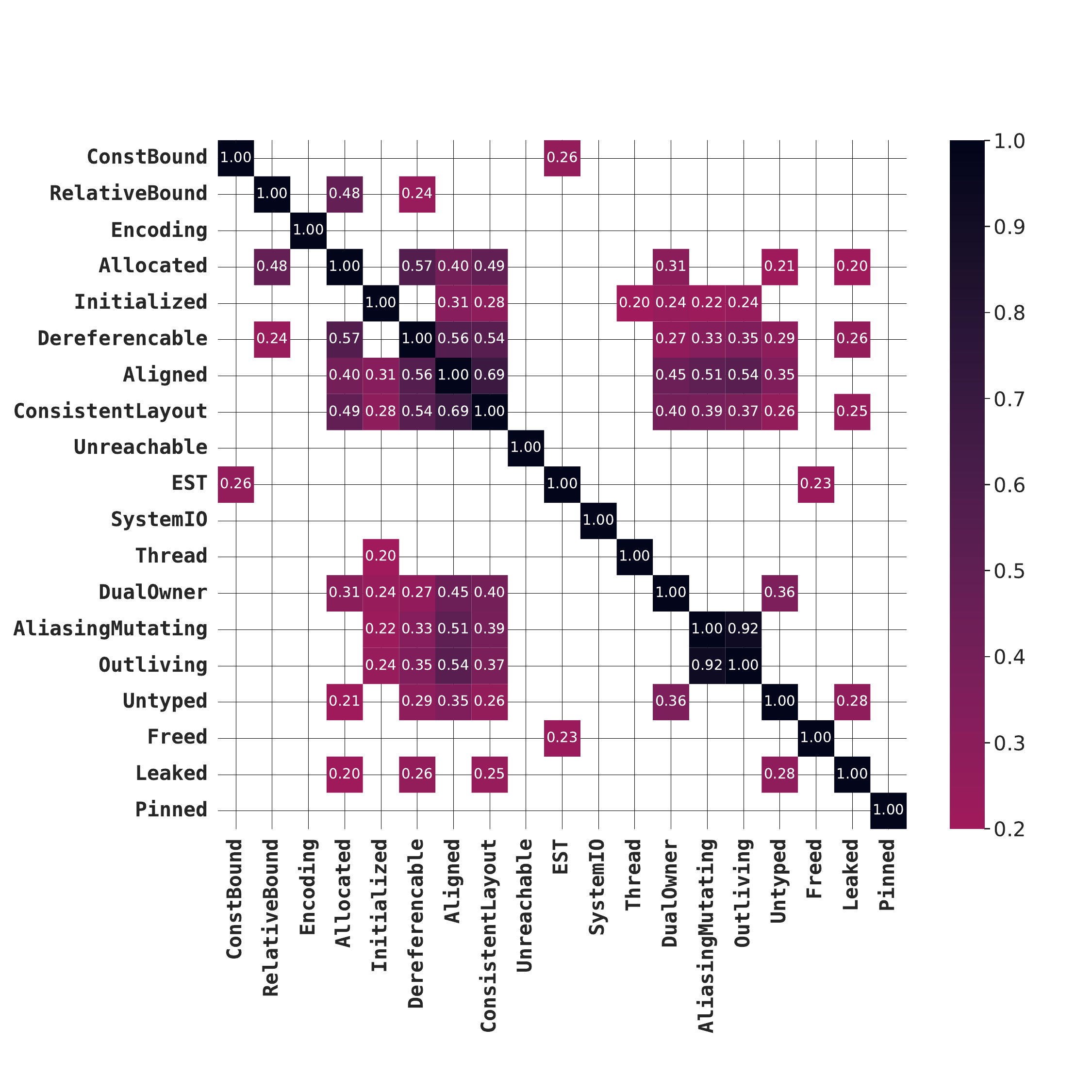}
	\caption{Correlation analysis results on the large dataset (original).}
	\label{fig:co1}
\end{subfigure}
\begin{subfigure}[]{0.49\textwidth}
	\includegraphics[width=\textwidth]{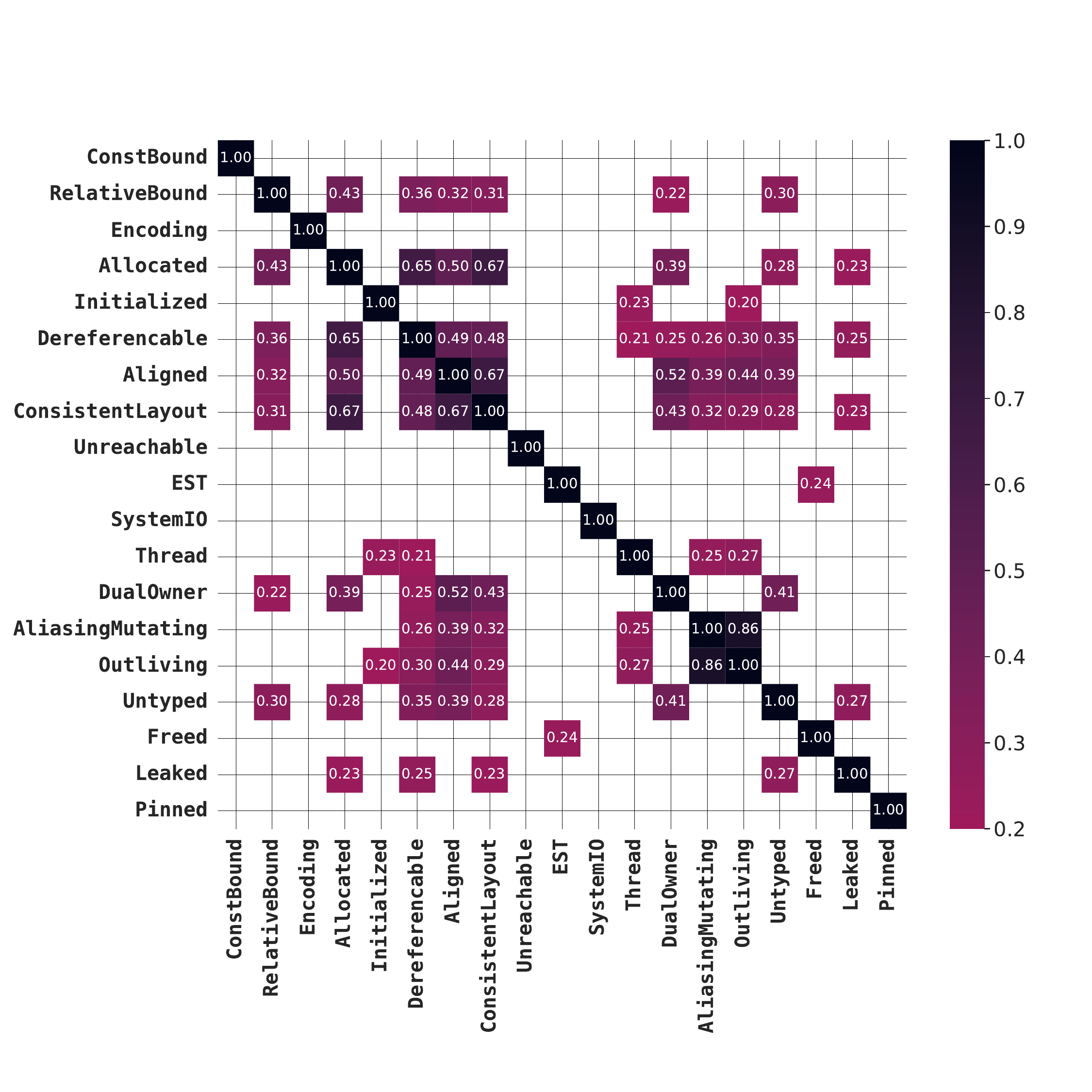}
	\caption{Correlation analysis results on the small dataset (filtered).}
	\label{fig:co2}
\end{subfigure}
\caption{Correlation matrices for both the large and small datasets. Each figure only includes the sections with weak correlation and above (correlation greater than 0.2).}
\label{fig:co}
\end{figure*}

\begin{goal}\label{g:gen}
	\textbf{Generality}: SP abstracts safety requirements not specific to one particular API's intricacies.
\end{goal}

\begin{goal}\label{g:amb}
	\textbf{Unambiguous}: SP intends to adopt the existing terminology and explanations as much as feasible in Rust.
\end{goal}

\begin{goal}\label{g:olp}
	\textbf{Nonoverlapping}: SP does not overlap, although they may be correlated.
\end{goal}

\begin{goal}\label{g:com}
	\textbf{Composability}: An Unsafe API's safety requirements can comprise several SPs.
\end{goal}

\begin{goal}\label{g:ess}
	\textbf{Essentiality}: Failure to comply with any SP would cause undefined behavior or additional vulnerabilities.
\end{goal}

\begin{goal}\label{g:prac}
	\textbf{Practicality}: SP is valuable and needs to be seriously considered in real-world programming scenarios.
\end{goal}

\begin{goal}
	\textbf{Unilingual}: SP disregards the Foreign Function Interface (FFI) and the intrinsic requirements of other programming languages.
\end{goal}

By adhering to these principles, the extracted safety properties aim to provide a comprehensive and practical understanding of the safety considerations associated with Unsafe Rust. It maintains compatibility with Rust's existing terminology and avoids unnecessary complexities related to FFI.

\subsubsection{Preprocessing} We noticed the redundancy in the standard library, such as \texttt{std} and \texttt{core} having an intersection. Thus we performed the following preprocessing for all unsafe APIs within \texttt{std}/\texttt{core}/\texttt{alloc} in Rust 1.70, including stable and nightly channels:

\begin{filter}
	For the methods exposed by both \texttt{core} and \texttt{std}, we kept only one of them.
\end{filter}

\begin{filter}\label{filter:num}
	For methods belonging to similar numeric types, we kept only one implementation.
\end{filter}

\begin{filter}
	For compiler intrinsics, we retained only those with no stable counterpart.
\end{filter}

As a result, we obtained a collection of unsafe APIs comprising 416 unsafe methods, with 127 being folded as 11 unique APIs by Filter~\ref{filter:num} (\textit{e.g.,} \texttt{unchecked\_mul::<u8>/::<u16>}~\cite{mulu8,mulu16} are merged). By applying the preprocessing step, we aimed to streamline and consolidate an unsafe API collection for further analysis and investigation.

\subsection{What Safety Properties Should We Satisfy?}
A code audit of all API documents within the collection was conducted to determine what safety properties correspond with the design goals. As shown in Table~\ref{table:sp}, we divided all safety properties into two main categories with 19 subdivisions.

\subsubsection{Working Procedure}
We simultaneously studied documents, defined safety properties, and labeled APIs. Regarding methodology, we performed two rounds of audits, with double-checking from the first and second authors. During the audit of each API, we focused on five sections: the functionality description, the safety description, the subchapters (including outer sections), the example code snippets, and the source code with its comments.

\underline{\textit{First Round: Initial Establishment.}} We labeled each std unsafe API with SPs that previously existed. The initial set of safety properties was empty. A new SP item was established if a safety requirement emerged and was not recorded in the current set. Therefore, any newly identified SP should be introduced for the first time during the first-round audit. If there are overlapped SPs, we merged them and re-checked related APIs to determine if they could be consolidated.

\underline{\textit{Second Round: Cross Checking.}} We observed that the API descriptions exhibited variations, including inconsistency and insufficiency. In the first round, 19 SPs were finally defined. In the second round, we focused on cross-checking and ensured that all unsafe APIs were appropriately labeled. We have paid particular attention to identifying any missing SP labels for each API that were not initially captured in the first round.

In summary, the first round of auditing ensured the completeness of SP categorization, while the second round enhanced the completeness of the required SP labels for each unsafe API.

\subsubsection{Categories}
We have divided the safety properties into two categories based on the state of the function execution as in program testing~\cite{berdine2005symbolic}, with no overlap between the sub-items.

\underline{\textit{Precondition Safety Property (PRE-SP).}} The precondition assumes that if the input values do not satisfy the safety requirements, the function call will trigger undefined behavior or additional vulnerabilities in Section~\ref{sec:ub}. Thus, any given API can be regarded as a black box for single-step execution~\cite{van1999silicon} at the call site, regardless of its internal implementation. PRE-SP complies with the initial characteristic of unsafe function (\textit{i.e.}, it cannot ensure safety for arbitrary inputs). In Table~\ref{table:sp}, we summarize 12 PRE-SPs. For example, \texttt{swap}~\cite{swap} has the description \textit{"Both x and y must be properly aligned."}, thus categorized into \texttt{Aligned}. 

\underline{\textit{Postcondition Safety Property (POS-SP).}} The previous assumption leads to the deduction that the function can be safely called if the proper inputs are supplied. However, this assurance only concerns the current program point. POS-SP focuses on the potential safety issues that may arise from the subsequent operations, assuming that the input values satisfy all PRE-SPs needed. In Table~\ref{table:sp}, we finally summarize 7 POS-SPs. For example, \texttt{zeroed}~\cite{zeored} has the description \textit{"There is no guarantee that an all-zero byte-pattern represents a valid value of some type T."}, thus categorized into \texttt{Untyped}.

The PRE-SP items are not nonoverlapping within POS-SPs through this separation. It can be verified by a Rust design, where creating raw pointers is always safe, but dereferencing them is unsafe~\cite{reference}. Similarly, we assume that PRE-SPs only affect the safety of function calls, while POS-SPs focus on the subsequent usage of inputs and return values. Furthermore, POS-SPs are only considered under the premise that all PRE-SPs are satisfied. Specifically, we merged \texttt{Aliasing} and \texttt{Mutating} based on the ground truth that all relevant APIs shared the same labels in these items. We empirically inferred that the primary side effect of breaking \texttt{Aliasing} rules is erroneously \texttt{Mutating} immutable data. We empirically inferred that the primary side effect of breaking \texttt{Aliasing} rules is erroneously \texttt{Mutating} immutable data.

\subsection{Correlation Analysis on Safety Properties}
We obtained a valuable dataset after completing the labeling for unsafe API collection. Although one of our design goals focuses on nonoverlapping, it is still necessary to investigate potential correlations between different SPs. This notion is from the empirical intuition that data satisfying \texttt{Dereferenceable} should always meet \texttt{Allocated}.

\subsubsection{Methodology} We conducted a correlation analysis based on two datasets, and their results demonstrate the anticipated differences. We will explain their characteristics first and then discuss the results of both datasets.

\underline{\textit{Large Dataset.}} The large dataset directly uses the original collection with labeled data, which includes the entire set of unsafe APIs. The labels of functionally related APIs may be similar. The intent of keeping a large dataset is to emphasize the quantity, as having adequate data can expose potential correlations.

\underline{\textit{Small Dataset.}} The small dataset is created by applying additional filter (Filter~\ref{filter:small}) to the large dataset. This is done to counteract the potential bias from excessive similar APIs. The small dataset intends to eliminate the redundancy of potentially irrelevant data and concentrate on diversity.

\begin{filter}\label{filter:small}
APIs must have the same labels and satisfy at least one of the following requirements:
\begin{enumerate}[0]
\item[$\bullet$] \noindent Implementations of the same method with different mutability.
\item[$\bullet$] \noindent Implementations of the same trait for different types, including mono-morphizations in the trait or struct definitions.
\item[$\bullet$] \noindent Functions with the same name implemented for different types within the same namespace.
\item[$\bullet$] \noindent Encapsulation of intrinsic functions.
\end{enumerate}
\end{filter}

\subsubsection{Correlation Matrix} As depicted in Figure~\ref{fig:co}, we built correlation matrices for both the large and small datasets, retaining only the elements with moderate correlation and above (correlation coefficient > 0.20). The large dataset has a higher susceptibility to interference from redundant APIs. For example, there are 30 implementations of the trait \texttt{SliceIndex<[T]>}~\cite{sliceindex}, all of which are labeled with \texttt{RelativeBound} and \texttt{Allocated} only. The correlation between them is thus higher in the large dataset, changing from 0.43 to 0.48. In the small dataset, the diverse functionality among APIs is more likely to result in the loss of pertinent data that could affect correlations. For example, the small dataset's correlation between \texttt{Aligned} and \texttt{AliasingMutating} decreases from 0.51 to 0.39. At last, \texttt{Encoding}, \texttt{Unreachable}, \texttt{SystemIO}, and \texttt{Pinned} achieve the best independence, as they have no substantial correlation with any other SPs in both matrices.

\begin{table}[]
\caption{SP pairs with correlation coefficients (CC) greater than 0.4 both in the large and small dataset correlation matrices.}
\label{table:pair}
\resizebox{\linewidth}{!}{
\begin{tabular}{|cc|>{\columncolor{gray_item}}c|cc|>{\columncolor{gray_item}}c|}
    \toprule[1pt]
	\textbf{SP1} & \textbf{SP2} & \textbf{AVG-CC} & \textbf{SP1} & \textbf{SP2} & \textbf{AVG-CC} \\
	\hline
	\rowcolor{gray_blank} \multicolumn{6}{|c|}{\textbf{Precondintion Safety Properties ONLY}}\\
	\hline
	Allocated       & RelativeBound    & 0.455 & Allocated       & Dereferencable   & 0.615 \\
	\hline
	Allocated       & ConsistantLayout & 0.580 & Allocated       & Aligned          & 0.450 \\
	\hline
	Dereferencable  & Aligned          & 0.525 & Dereferencable  & ConsistantLayout & 0.510 \\
	\hline
	Aligned         & ConsistantLayout & 0.685 & \multicolumn{3}{c|}{} \\
	\hline
	\rowcolor{gray_blank} \multicolumn{6}{|c|}{\textbf{Precondintion Safety Properties with Postcondition Safety Properties}}\\
	\hline
	DualOwner       & Aligned          & 0.485 & DualOwner       & ConsistantLayout & 0.415 \\
	\hline
	Outliving       & Aligned          & 0.490 & Outliving       & AliasingMutating & 0.895 \\
    \bottomrule[1pt]
\end{tabular}
}
\end{table}

\underline{\textit{Case Study.}} Based on two diagrams in Figure~\ref{fig:co}, we extracted all the pairs with at least a moderate CC, as listed in Table~\ref{table:pair}. Among the six pairs with no POS-SPs, we empirically found that they are related to dereferencing operations. Although dereferencing was not considered a distinct item in the SP category, such operations are pervasive in the inner code of unsafe methods. We inferred knowledge about safety requirements for dereferencing that was not explicitly categorized: The first-class SP for a \textit{valid pointer} with the highest priority is \texttt{Allocated}, followed by \texttt{Dereferencable}, \texttt{ConsistentLayout}, and \texttt{Aligned}. \texttt{RelativeBound} should be considered if it has pointer arithmetic. Even though \texttt{Initialized} is not stated in Table~\ref{table:pair}, we still view it as a prerequisite for dereferencing, as Rust's undefined behavior has explicit requirements for valid values of raw pointers. In this paper, we advocate for the safe usage of raw pointers by satisfying these 6 PRE-SPs.

\section{Verifying Real-world Unsafe Code}\label{sec:cveio}
This section presents the practical usability of our classification in real-world scenarios and the frequency of unsafe API usage in the Rust ecosystem. We classify existing CVEs~\cite{cve} to validate SP coverage and conduct a statistical analysis of unsafe API usage on crates.io~\cite{cratesio}.

\subsection{Classifying Existing Unsafe CVEs}\label{sec:cve}

\subsubsection{Workflow} The workflow consists of two primary steps: Create a database of CVEs caused by misusing unsafe cust and classify them into safety properties by manual code review.

\underline{\textit{CVE Set.}} We employed the CVE dataset from the CVE program (https://cve.mitre.org) and searched on the CVE list using the keyword \textit{\textbf{"Rust"}}. The results are sorted by CVE ID in chronological order (\textit{i.e.,} submission date). The final CVE dataset ranged from CVE-2017-20004~\cite{cve20004} to CVE-2023-30624~\cite{cve30624}. We initially filtered CVE based on CVE descriptions, primarily retaining memory-safety issues. Unrelated CVEs were removed, such as leaking sensitive information. We filtered those CVEs triggered in the panic path because this study does not specifically work for panic safety~\cite{rustonomicon}. Additionally, the retained CVEs cannot be located in a deprecated or yanked crate and should have a link to the source code to support a code audit.

\begin{figure}[]
\begin{subfigure}[]{0.45\textwidth}
\begin{lstlisting}[language=Rust, style=colouredRust, label=list:cvecode, caption=Source code of CVE-2021-45709 in the crypto2 crate through 2021-10-08 for Rust.\\]
#[inline]
fn xor_si512_inplace(a: &mut [u8], b: &[u32; Chacha20::STATE_LEN]) {
    unsafe {
        let d1 = core::slice::from_raw_parts_mut(a.as_mut_ptr() as *mut u32, Chacha20::STATE_LEN);
        for i in 0..Chacha20::STATE_LEN {
            d1[i] ^= b[i];
        }
    }
}
\end{lstlisting}
\label{list:cvecode}
\end{subfigure}

\begin{subfigure}[]{0.45\textwidth}
\begin{lstlisting}[language=c, style=colouredRust, label=list:cvedoc, caption=Description documented in RUSTSEC-2021-0121.\\]
Description
The implementation does not enforce alignment requirements 
on input slices while incorrectly assuming 4-byte alignment 
through an unsafe call to std::slice::from_raw_parts_mut, 
which breaks the contract and introduces undefined behavior.
\end{lstlisting}
\label{list:cvedesp}
\end{subfigure}
\caption{Example of the CVE classification. The buggy source code and description of CVE-2021-45709 (in RUSTSEC). This CVE violates the safety requirement of \texttt{Aligned} and triggers UB when using the unsafe API slice::from\_raw\_parts\_mut.}
\label{fig:cveexample}
\end{figure}

\begin{figure}[]
\includegraphics[width=0.5\textwidth]{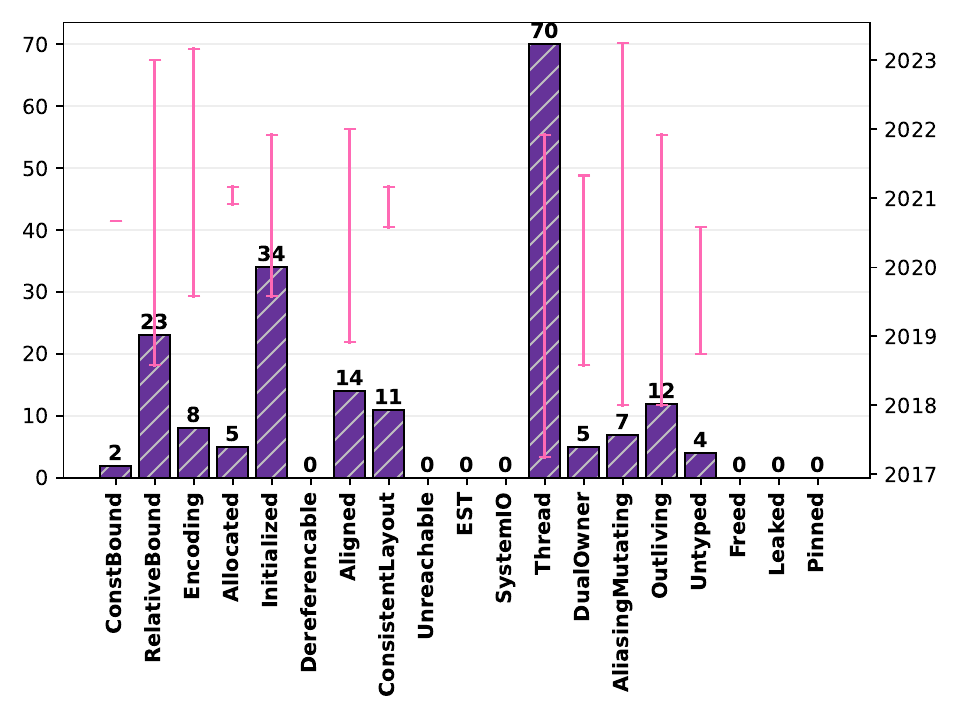}
\caption{SP classification results and their duration on existing Rust CVEs. These CVEs are memory-safety issues resulting from misusing unsafe code and ignoring unrelated issues, such as leaking sensitive information.}
\label{fig:cve}
\end{figure}

\underline{\textit{CVE Audit.}} We performed a manual audit of error snippets that led to security issues. The first and second authors double-checked the results. We painstakingly investigated whether misusing unsafe code was the root cause of each CVE. Due to the short descriptions provided on the CVE website, we utilized various sources, including issues, pull requests, contributors (\textit{e.g.,} RUSTSEC~\cite{rustsec}), and fixed code, to pinpoint the source code related to each CVE. Any CVEs that did not satisfy the front criteria were removed from our dataset. Around 86.73\% of the 196 CVEs in the final dataset were attributed to misusing unsafe APIs from the standard library. In contrast, the remaining CVEs were caused by dereferencing raw pointers, using non-std unsafe functions, or outside FFI. Based on the descriptions and code reviews, we further classified each CVE into the SPs it violated.

\underline{\textit{CVE Example.}} Figure~\ref{fig:cveexample} presents an example of a classified CVE derived from CVE-2021-45709~\cite{cve45709,sec0121}. Based on the explicit error description in its issue, we were able to locate the buggy source code and confirm incorrect usage of \texttt{from\_raw\_parts\_mut}~\cite{fromptrmut}. This API was annotated by the following SPs: ConstBound, RelativeBound, Allocated, Dereferencable, Aligned, ConsistantLayout, AliasingMutating, and Outliving. This CVE violates the requirement of \texttt{Aligned}, leading to undefined behavior.

\subsubsection{Results and Benchmark} We conducted a study on 404 CVE descriptions and performed a code review on the remaining 196 CVEs after filtering. We classified them based on SP categorization and analyzed their distribution. It has been manually verified that the causes of these CVEs do not exceed our SP classification. Finally, we generated a benchmark encompassing the various SPs in the identified CVEs.

\underline{\textit{SP and Time-span Distribution.}} Figure~\ref{fig:cve} depicts the classification results. \texttt{RelativeBound}, \texttt{Initialized}, and \texttt{Thread} had a significant number of CVEs (at least 23). Following them, there are fewer CVEs associated with \texttt{Aligned}, \texttt{Outliving}, \texttt{ConsistentLayout}, and the other 6 SPs (ranging from 2 to 14). 7 SPs have no corresponding CVEs. Except for \texttt{ConsistentLayout}, \texttt{ConstBound}, and \texttt{Allocated}, the time span of CVEs for each SP ranges from as early as August 2019 to as late as December 2021 from a temporal perspective. This period contains approximately 91.84\% of listed CVEs.

\begin{table}[]
\caption{Open-source code analyzers that can detect specific SPs. Each static bug detection tool may not necessarily support all scenarios related to the corresponding SP.}
\label{table:tool}
\resizebox{\linewidth}{!}{
\begin{tabular}{cp{6.8cm}}
    \toprule[1pt]
	\textbf{Tool} & \textbf{Supported SPs of Each Static Analyzer}\\
	\midrule[1pt]
	\textbf{\textsc{Rudra}}~\cite{bae2021rudra} & \texttt{Thread}, \texttt{DualOwner}, \texttt{Initialized}\\
	\midrule[1pt]
	\textbf{\textsc{SafeDrop}}~\cite{cui2023safedrop} & \texttt{Allocated}, \texttt{DualOwner}, \texttt{Freed}, \texttt{Initialized}\\
	\midrule[1pt]
	\multirow{2}{*}{\textbf{\textsc{MirChecker}}~\cite{li2021mirchecker}} & \texttt{Const-Numeric Bound}, \texttt{Relative-Numeric Bound}, \texttt{DualOwner}\\
	\midrule[1pt]
	\textbf{\textsc{FFIChecker}}~\cite{li2022detecting} & \texttt{Allocated}, \texttt{Leaked} \textit{(Rust/C FFI Only, based on LLVM)} \\
	\midrule[1pt]
	\textbf{\textsc{rCanary}}~\cite{rCanary} & \texttt{Leaked} \\
    \bottomrule[1pt]
\end{tabular}
}
\end{table}  

\underline{\textit{Benchmark on SP.}} Our classification results provide a set of CVE lists for each SP that can be used as a benchmark. This benchmark can be employed to evaluate the effectiveness of research prototypes or bug-detection tools designed for particular SPs. As far as we know, the Rust community needs a unified, ground-truth-supported benchmark to support effectiveness comparisons based on safety issues. We advocate for setting such a benchmark to serve as a basis for the community. We also list some open-source code detection tools for Rust programs, all of which serve for specific SPs, as shown in Table~\ref{table:tool}.

\underline{\textit{Case Study.}} The most CVEs were caused by \texttt{Thread} (70) violations. This is predominantly the result of user-defined types that unconditionally implement the \texttt{Send}~\cite{send}/\texttt{Sync}~\cite{sync} traits or fail to ensure that \texttt{Send}/\texttt{Sync} implementations have the correct bounds. Such violations may result in data races and memory-safety issues across the thread boundaries. \texttt{Initialized} (37) was the second most common SP, and its typical scenarios are as follows: 1) Create an uninitialized buffer and pass it to the user-defined \texttt{Read}~\cite{read} implementation, allowing safe code to read uninitialized memory; 2) Increase buffer length without reserving memory, causing write-out-of-bound or dropping uninitialized memory issues; 3) Create an uninitialized \texttt{NonNull} pointer. 

\underline{\textit{Discussion.}} We observed that the statistic in Figure~\ref{fig:cve} may not accurately reflect the frequency of SP usage in the real world. For example, a significant portion of CVEs on \texttt{Initialized} (85.3\%) and \texttt{Thread} (92.9\%) were discovered by the \texttt{sslab-gatech} using their static analyzer \textsc{Rudra}~\cite{bae2021rudra}, which is designed to detect bugs inspired by specific bug patterns. This observation suggests that analyzers designed for bug patterns may effectively identify vulnerabilities that violate specific SPs. On the other hand, many undiscovered vulnerabilities related to specific SP may not have been registered on https://cve.mitre.org.

\subsection{Statistics on crates.io Ecosystem}\label{sec:cratesio}
The findings from Section~\ref{sec:cve} indicate that 86.73\% of the classified CVEs were caused by misusing unsafe APIs in std. This observation prompted us to conduct a statistical analysis of the usage of unsafe APIs within the Rust ecosystem.

\begin{figure}[]
\includegraphics[width=0.5\textwidth]{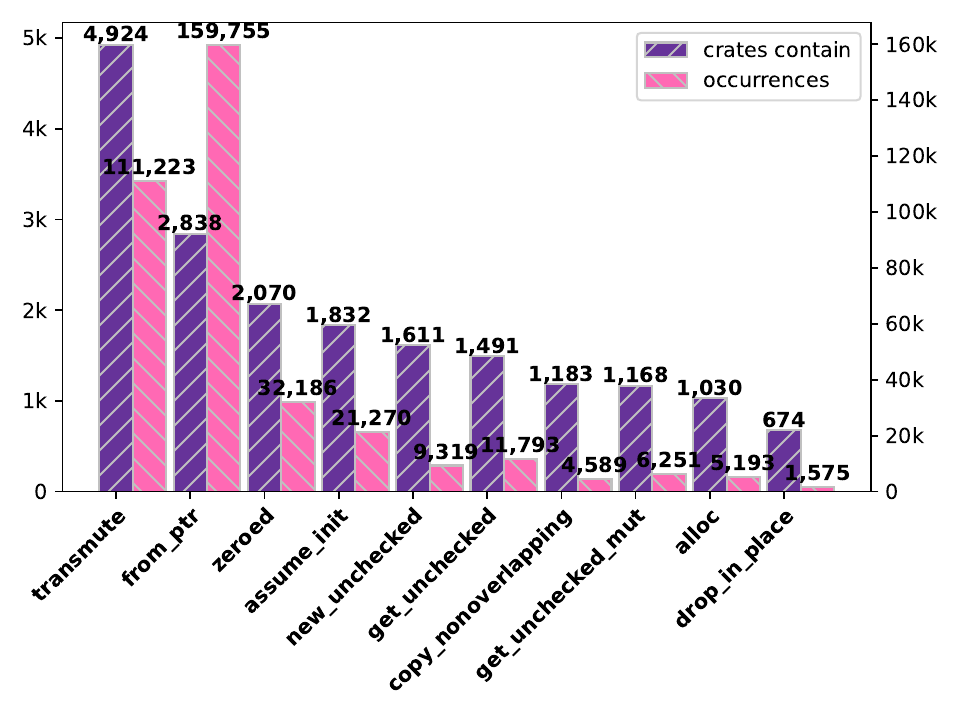}
\caption{Statistics results on unfiltered strings (unsafe APIs) in the crates.io ecosystem. The top ten most frequently used strings across all repositories and their source code occurrences sorted by the sum of crates.}
\label{fig:crates.io}
\end{figure}

\subsubsection{Open-source Crates Database} As crates.io is the crate management platform in the Rust community, we used all its repositories to serve as a code database. To evaluate the frequency of unsafe API usage, we matched regular expressions to source code without compilation. Using function name as the criterion, we merged identical unsafe APIs, creating a dictionary of 140 unique unsafe API strings. Then we removed 20 strings that have the same name as other safe functions in the std (\textit{e.g.}, \texttt{add}~\cite{addsafe,addunsafe}). To improve the accuracy of the statistics, we only included per-file instances if the \textit{unsafe} keyword was used in the source code.

\subsubsection{Frequency Statistics} As of 2023-01-30, we mined all the latest crates from crates.io. The statistic indicates that 21,506 crates use Unsafe Rust among the 103,516 crates on crates.io (3,614 are yanked). For each string, we collected a statistical summary, including the number of crates in which the string appears and the total usage count of the string across all crates. The top ten most frequently used strings are listed in Figure~\ref{fig:crates.io}, which depicts statistical results in two dimensions. We observed that the primary scenarios encompass type conversions (\texttt{transmute}), manual memory management (\texttt{zeroed}, \texttt{alloc}, \texttt{drop\_in\_place}), unsafe constructors (\texttt{new\_unchecked}), deferred initialization (\texttt{assume\_init}), unsafe indexing (\texttt{get\_unchecked/mut}), unsafe referencing (\texttt{from\_ptr}), and unsafe memory copies (\texttt{copy\_nonoverlapping}). Note that the results presented above do not account for filtered strings (\textit{e.g.,} \texttt{read}, \texttt{as\_mut}, \texttt{from\_raw}, etc.).

\section{Surveying Rust Programmers}\label{sec:survey}
In this section, we conducted an online survey on Goldendata~\cite{golendata} to evaluate the Safety Properties in precision, significance, usability, and frequency from the perspective of experienced Rust developers.

\subsection{Methodology}

\subsubsection{Recruitment} We require participants to be at least 18 years old with a minimum of 1 year of experience in Rust programming and to have written at least 5,000 lines of Rust code, including over 1,000 lines of unsafe code. We posted our survey on the Rust-related community to recruit volunteers and emailed contributors from Rust-lang and the popular repositories on crates.io.

\subsubsection{Procedure} We provide each defined SP with a representative unsafe API for participants on each page. Note that the relationship between API and SP is many-to-many. Hence, the given API only targets one SP. For each API, we further supply a triplet $(S,\ P_1,\ P_2)$, containing a document slice $S$ for the current SP, a sound code snippet $P_1$, and a misused PoC $P_2$. $P_1$ and $P_2$ are carefully designed to be short and easy to debug. $P_2$ violates the safety requirements of the current SP, ensuring that all other SPs are satisfied. We link $S$ to the online document for referencing, and both $P_1$ and $P_2$ can be redirected to the Rust Playground~\cite{playground} for online execution. Furthermore, 18 out of 19 $P_2$ will trigger UB that can be captured by \textsc{Miri}~\cite{miri}, making it easier for participants to understand issues.

Participants must read the definition and the triplet of each SP. Then we ask them four questions listed below:

\begin{enumerate}[0]
\item[$\bullet$] \noindent \textbf{Q1:} We asked participants to rank the accuracy of SP definitions. Can the safety requirements of each SP be explained in a concise and precise definition?
\item[$\bullet$] \noindent \textbf{Q2:} We asked participants to rank the significance of each SP. Does violating each SP lead to unacceptable issues?  Is it necessary to document such requirements explicitly in \texttt{Rustdoc}?
\item[$\bullet$] \noindent \textbf{Q3:} We asked participants to rank the usability of each SP. Should users consider the context of SP satisfaction in real-world unsafe programming? Does adhering to each SP help write sound code?
\item[$\bullet$] \noindent \textbf{Q4:} We asked participants to rank the frequency of each SP. How frequently do they encounter situations that require careful use of this SP? Is it often considered when crossing unsafe boundaries?
\end{enumerate}

For each of the four questions, we have devised a $[-1, 1]$ scoring scale to represent negative, neutral, and positive responses. It is crucial to note that these four questions have no objectively correct answers. The participants' responses may vary based on their unique perspectives and experiences.

\subsection{Survey Results} We distributed the survey between July 10 and July 25, 2023, and received 90 responses. After review by the first and second authors, it was determined that 50 were valid. The criteria for a valid response included excluding surveys with excessively short completion times (less than 15 minutes), the same pattern throughout the entire survey, and responses inconsistent with Rust facts.

\begin{figure}[]
\includegraphics[width=0.49\textwidth]{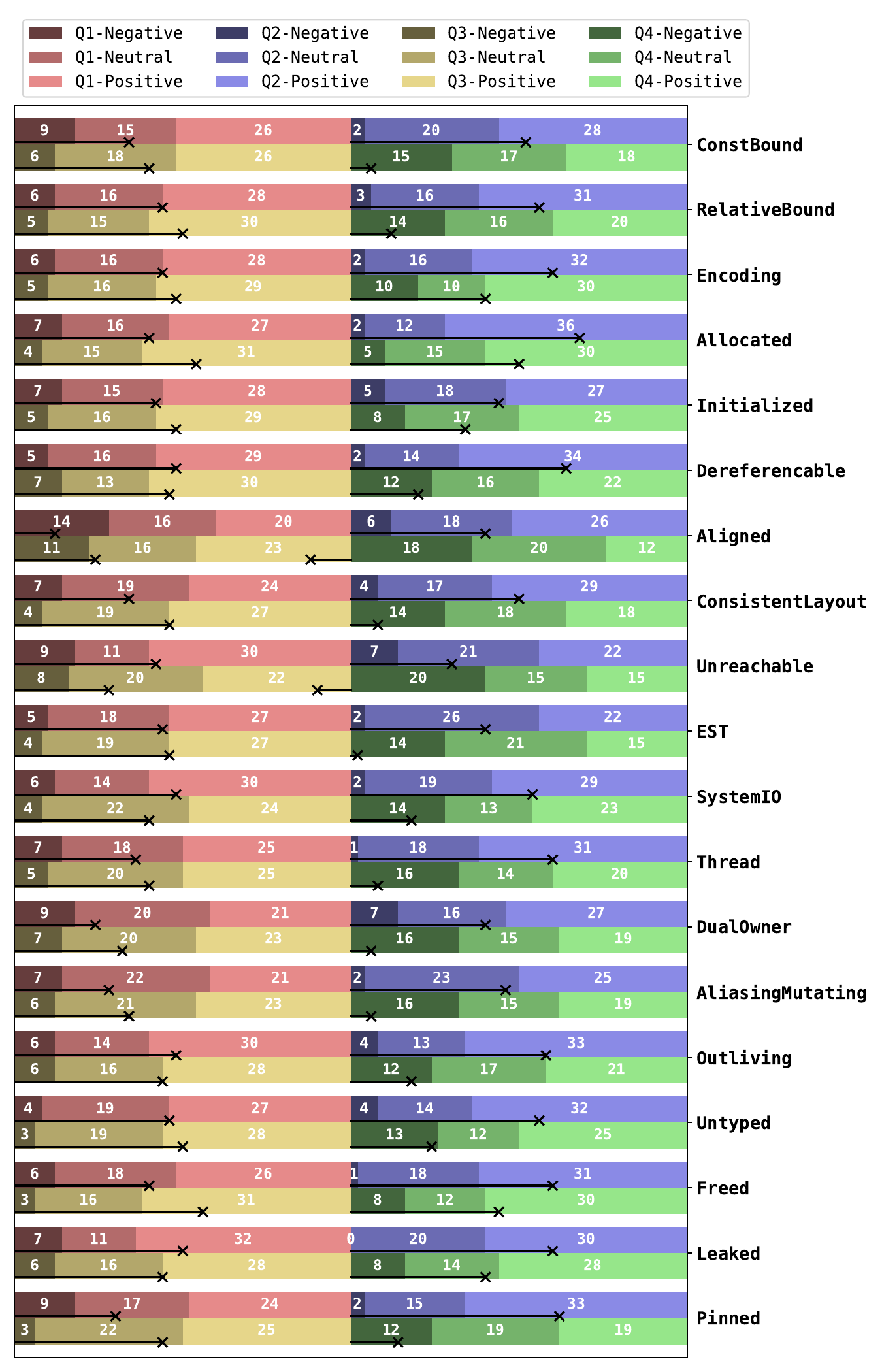}
\caption{Survey results on Unsafe Rust programmers. Precision, significance, usability, and frequency are the four dimensions rated for each SP. Positivity, neutrality, and negativity are the options for each dimension.}
\label{fig:survey}
\end{figure}

\underline{\textit{Q1: Precision Ratings.}} The results of Q1 illustrate participants' comprehension and endorsement of SP definitions, providing an in-depth appraisal of \cref{g:gen,g:amb,g:olp,g:com}. The score distribution is $[6,25]$, with a mean of 19.3, a standard deviation of 4.8, and 16 SPs scored greater than 15. We observed that SPs with brief descriptions tended to receive higher scores such as \texttt{SystemIO} (24) and \texttt{Leaked} (25). Whereas SPs with subcategories had a lower score due to the comprehension threshold, such as \texttt{ConsistentLayout} (17) and \texttt{AliasingMutating} (14). Participants exhibited a comparatively negative attitude toward unusual SPs, specifically \texttt{Aligned} (6). They are also unclear about the newly introduced categories like \texttt{DualOwner} (12). In the optional supplementary comments, 4 participants emphasize the necessity for the document to highlight the side effects caused by violating \texttt{DualOwner}.

\underline{\textit{Q2: Significance Ratings.}} The results of Q2 indicate their perspectives on the safety issues caused by violating each SP, in accordance with \cref{g:ess}. The score distribution is $[15,34]$, with a mean of 26.3, a standard deviation of 5.0, and 18 SPs scored greater than 20. We noticed that participants tend to be positive concerning memory safety. Rust developers devote particular attention to memory safety and evince a predictable sensitivity to the safety requirements of unsafe code. However, an exception existed, which is \texttt{Unreachable} (15). The majority of participants viewed \texttt{Unreachable} as an inconsequential problem and instinctively assumed that panic is always memory-safe, losing focus on potential threats to panic safety.

\underline{\textit{Q3: Usability Ratings.}} The results of Q3 represent the sensitivity to the context when crossing unsafe boundaries in real-world unsafe programming, thus addressing \cref{g:prac}. The score distribution is $[12,28]$, with a mean of 21.4, a standard deviation of 4.2, and 15 SPs scored greater than 20. Notably, there is a significant correlation between the scores in Q2 and Q3. It can be explained by assuming that users may be less likely to check the requirements in real-world situations if they perceive one SP as insignificant. Conversely, if their programs are less affected by one SP in practice, they may perceive it as unimportant, which is consistent with their intuition. We observed that 89.4\% of the Q2 scores are higher than Q3, indicating that participants may have a higher awareness of significance than programming habits in real-world practice.

\underline{\textit{Q4: Frequency Ratings.}} The results of Q4 reveal the frequency with which Rust developers encounter each SP in unsafe programming. The score distribution is $[-6,25]$, with a mean of 8.6 and a standard deviation of 8.8. These responses further validated the results in Section~\ref{sec:cratesio}, which measured the frequency of unsafe API usage in crates.io. We found significant discrepancies in the score distribution for this question. It shows that \texttt{Allocated} (25), \texttt{Freed} (22), \texttt{Leaked} (20), and \texttt{Initialized} (17) are encountered more frequently in Unsafe Rust. We infer that these SPs are tightly connected with common scenarios, including manual memory management and deferred initialization. As for \texttt{Encoding} (20), users may use it frequently to interact with C code in unsafe contexts. For SPs with lower or even negative scores, we suggest that the Rust developers and community may need to pay more attention to avoid misusing them.

The survey results confirmed our classification of safety properties with statistical significance. It is necessary to define a systemic classification, as experienced Rust programmers highly care about memory safety issues caused by unsafe code. At last, it also reveals a significant variation in the occurrence frequency of different SPs in real-world Rust programs.

\section{Threats to Validity}
For internal validity threats, using std unsafe documents as our knowledge base might not provide exhaustive coverage for investigating the categories toward security requirements. We adopted a validation based on the CVE classification to address this limitation. Also, survey participants might not be representative enough; some might be malicious respondents, cheat on the programming experience, or send multiple submissions. To ensure internal validity, various measures were implemented. First, we utilized multiple recruitment channels, such as private email invitations. Second, we clearly outlined the mandatory requirements for programming experience. Third, the first and second authors manually verified all the responses. Fourth, we imposed restrictions on the number of submissions from the same IP address.

For external validity threats, due to the ongoing development of Rust, the programming style and usage of Unsafe Rust may evolve over time. Despite considering both stable and nightly channels, future updates may introduce new unsafe APIs, modify API descriptions and implementations, or even deprecate some unsafe APIs. We also acknowledge the existence of uncommon safety descriptions that cannot be classified based solely on the std unsafe documents or existing CVEs because they have not been documented in any std unsafe document or existing CVE.

\section{Related Work}
\textbf{\textit{Empirical Studies on Rust Security.}} Researchers have conducted empirical studies to understand how to use Unsafe Rust from real-world Rust Programs~\cite{astrauskas2020programmers, evans2020rust, qin2020understanding, unsafesyntactic, yu2019fearless} and existing CVEs~\cite{xu2021memory}. They summarize valuable bug patterns and provide insights into different aspects of safety guarantees. However, these studies do not extract safety requirements from the safety descriptions in the API documents. Several empirical studies focus on the Rust learning curve~\cite{abtahi2020learning, fulton2021benefits} and the programming challenges introduced by compiler errors~\cite{zhu2022learning}. Researchers also leveraged Rust-related Stack Overflow data to understand real-world development problems~\cite{zhu2022learning}. However, we are more concerned with experienced system engineers who are proficient than Rust beginners. They need Unsafe Rust to achieve low-level control and better understand the safety requirements when crossing unsafe boundaries.

\noindent \textbf{\textit{Bug detection methods in Rust.}} As a strongly typed programming language, formal verifications for Rust have received considerable attention~\cite{lattuada2023verus, ho2022aeneas, astrauskas2022prusti, dang2019rustbelt, jung2019stacked, jung2017rustbelt, wolff2021modular, matsushita2021rusthorn, hahn2016rust2viper, crichton2022modular}. Existing studies also employ static or dynamic analysis to detect bugs in Rust programs, including symbolic execution~\cite{lindner2018no, mirai}, model checking~\cite{toman2015crust, vanhattum2022verifying, bae2021rudra, li2021mirchecker, cui2023safedrop}, interpreter~\cite{miri}, and fuzzing~\cite{dewey2015fuzzing, jiang2021rulf}. We have discovered that some of the above prototypes analyze errors based on bug patterns corresponding to our SP categorization. Table~\ref{table:tool} illustrates the relationship between the relevant tools and their supported SPs. At last, we believe this paper will encourage researchers and developers to focus on diverse safety requirements when handling unsafe boundaries. In the future, more testing tools can be developed to aid in the detection of safety violations for different SPs.

\section{Conclusion}
As Rust is a system programming language, Unsafe Rust is integral to achieving low-level control over implementation details. With increasing system software adopting Rust, understanding the safety requirements when crossing unsafe boundaries is crucial, particularly with well-defined categorization. To this end, we conducted the first comprehensive empirical study on safety requirements across the unsafe boundary. We focus on unsafe API documents in the standard library to infer safety properties, then categorize unsafe APIs and existing CVEs. Additionally, we conducted a user survey to gain insights into four aspects of these safety properties from experienced Rust developers. Through these efforts, we aim to promote the standardization of systematic documents for Unsafe Rust in the Rust community.

\bibliographystyle{ACM-Reference-Format}
\bibliography{unsafe.bib}

\end{document}